\documentclass[twocolumn,aps,prb,groupedaddress,showpacs,raggedbottom,nobalancelastpage,amssymb]{revtex4-1}

\usepackage{graphicx}
\usepackage{amsmath}
\usepackage{amssymb}
\usepackage{pgf}
\usepackage{bm}
\usepackage{amsfonts}
\usepackage{appendix}
\usepackage{dsfont}
\usepackage{hyperref}

\hypersetup{
  colorlinks = true, linkcolor = brown, citecolor = purple
}

\newcommand{\ssm}{\scriptscriptstyle\rm}

\newcommand{\overbar}[1]{\mkern 1.5mu\overline{\mkern-1.5mu#1\mkern-1.5mu}\mkern 1.5mu}

\begin{document}

\title{Single spin probe of Many-Body Localization}
\author{Evert P.L. van Nieuwenburg}
\affiliation{Institute for Theoretical Physics, ETH Zurich, 8093 Z{\"u}rich, Switzerland}
\author{Sebastian D. Huber}
\affiliation{Institute for Theoretical Physics, ETH Zurich, 8093 Z{\"u}rich, Switzerland}
\author{ R. Chitra}
\affiliation{Institute for Theoretical Physics, ETH Zurich, 8093 Z{\"u}rich, Switzerland}

\begin{abstract}
We use an external spin as a dynamical probe of many body localization. The probe spin is coupled to an interacting and disordered environment described by a Heisenberg spin chain in a random field. The spin-chain environment can be tuned between a thermalizing delocalized phase and non-thermalizing localized phase, both in its ground- and high-energy states. We study the decoherence of the probe spin when it couples to the environment prepared in three states: the groundstate, the infinite temperature state and a high energy N\'eel state. In the non-thermalizing many body localized regime, the coherence shows scaling behaviour in the disorder strength. The long-time dynamics of the probe spin shows a logarithmic dephasing in analogy with the logarithmic growth of entanglement entropy for a bi-partition of a many-body localized system. In summary, we show that decoherence of the probe spin provides clear signatures of many-body localization.
\end{abstract}


\maketitle

\section{Introduction}
The transition from the realm of quantum mechanics to the classical world is signaled by the loss of coherence. Understanding how this decoherence takes place is both a fundamental question as well as of relevance for technological applications that try to make use of quantum coherence. On the other hand, the way a well-controlled quantum system loses its coherence can provide important insights into the properties of the environment. In particular, one can see whether the environment constitutes a thermal bath \cite{Giraldi2013, Kof2004, Bre02a}. A recent intensely discussed issue is the extent to which a disordered many-body system can be considered thermal \cite{Nandkishore2015}. Or, in other words, if it can remain many body localized and hence non-ergodic at a finite energy density.

Such a many-body localized (MBL) system is an intriguing phase of matter. A key issue is that for a generic many-body state at finite energy density $\epsilon$, one might expect the eigenstate-thermalization-hypothesis (ETH) to hold, where expectation values of local operators exhibit a thermal behaviour at a temperature $T_{\ssm eff}\sim \epsilon$. However, it has now been established that this does not have to be the case~\cite{Basko2006, Pal2010, Vos2013, Serbyn2013, Kja2014,Bah15a, Nandkishore2015, Schreiber2015, Smith2015, Gopalakrishnan2016, Imbrie2016, Roe2016}. 

While the failure of the ETH is a fascinating phenomenon, it entirely relies on the perfect isolation of a quantum system from the environment: any globally accessible heat-bath will necessarily equilibrate the system and render the notion of ETH obsolete. In this respect, probing an MBL system represents a difficult experimental challenge, as any probe necessarily induces some degree of coupling to the outside world~\cite{Johri2015, Fischer2016, Nandkishore2014}. Regardless, recent experiments based on cold atoms~\cite{Schreiber2015,Choi16a} and trapped ions~\cite{Smith2015} have tackled this challenge. The experiments show that order in the initial state can be preserved for much longer timescales in the presence of disorder and interactions than in the case without, although the long term behaviour is governed by the coupling to the environment~\cite{Fischer2016,Schreiber2015}. All in all, there is a need for experimentally relevant probes of many-body localization~\cite{Moessner15,Eisert16}. 

In this paper, we propose to use the decoherence of a single spin coupled to a disordered spin-chain as a way of probing many-body localization~\cite{Serbyn2014,Abanin14b,Vasseur2015}. Our work can be seen as a minimal probe for such an MBL system, and does not assume weak coupling. 

The study of the decoherence dynamics of a single spin attached to a bath has a long history \cite{Cucchietti2005,Lag2005, Camalet2007, Rossini2007, Ou2007, Wu2014, Wu2016, Restrepo2013}. In particular, the roles of soft modes close to second order phase transition \cite{Wino2009, Camalet2007}, or the presence of a non-vanishing order-parameter in the bath have been studied in depth. Moreover, the relation of the level statistics of the bath to the decoherence properties of the single spin is well established \cite{Brox2012}. Here, we want to capitalize on these insights in order to use the decoherence of a single spin as a \emph{probe} for a many-body localized system.
\begin{center}
	\begin{figure}[b]
			\includegraphics[scale=0.8]{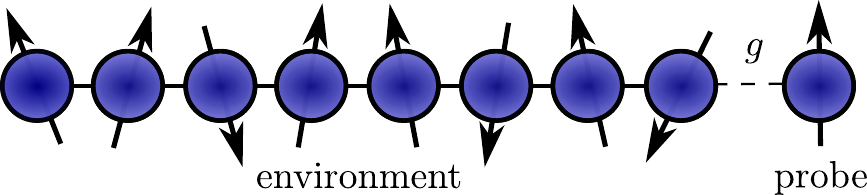}
			\caption{Sketch of the setup. A single probe spin is coupled to (the end of) a disordered and interacting spin-chain forming the environment. The probe is coupled in such a way that its dynamics is governed purely by dephasing, which is strongly influenced by the state of the environment. 
			\label{fig:system}}
	\end{figure}
\end{center} 

\vspace{-1.0cm}
\section{Model}
The MBL reservoir that we study is a Heisenberg spin chain of size $L$ in a random field, cf. Fig.~\ref{fig:system}: 
\begin{equation}
	\label{eq:environmentHamiltonian}
H_{\textrm{s}} = J \sum_{i=1}^{L-1} \boldsymbol{S}_i \cdot \boldsymbol{S}_{i+1} + \sum_{i=1}^L h_i S^z_i,
\end{equation}
where $J>0$ is the exchange coupling (which we fix to $J=1$ for the rest of this work), $\boldsymbol{S}$ are the canonical spin-$1/2$ operators and the random fields $h_i$ are independently and uniformly distributed in the interval $[-h_{\ssm max},h_{\ssm max}]$. This model is known to exhibit a transition between a many-body localized phase and a thermalizing phase as a function of $h_{\ssm max}$~\cite{Pal2010}. Let us review the key properties of this system in the light of its use as an effective bath for a probe spin.

Previous studies~\cite{Agarwal2015,Geraedts2016} indicate that in the thermodynamic limit the MBL transition occurs at $h_{\ssm max} \sim 3.5$. At values $h_{\ssm max} \lesssim 3.5$, the closed system is in a delocalized phase that obeys the ETH. For larger values the system is many-body localized, violates ETH and hence avoids thermalization. 

\begin{center}
	\begin{figure*}[ht!]
			\includegraphics{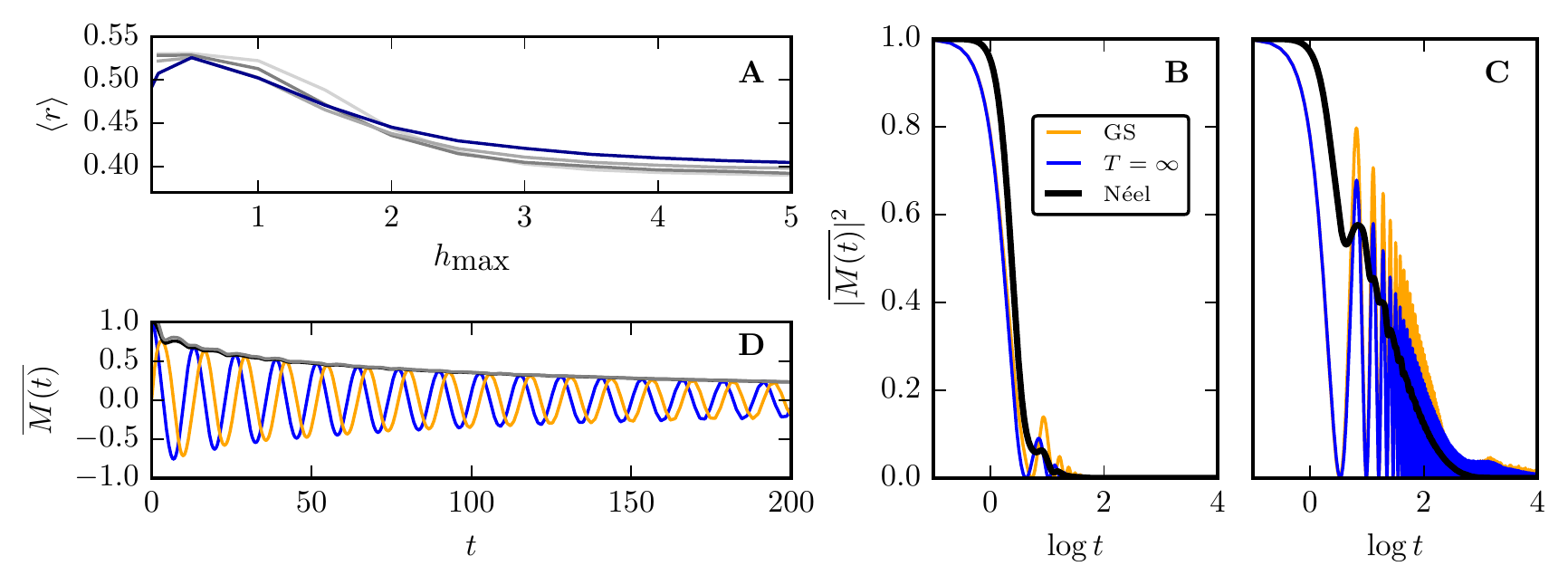}
			\caption{\textbf{A}: The level spacing statistics for $L=8$ (blue) up to $L=14$ (lightest grey) in steps of two, summarized by the $\langle r \rangle$ statistic~\cite{Oganesyan2007}. For a Wigner-Dyson distribution $\langle r \rangle$ approaches the value $\sim 0.53$, whereas for the Poissonian distribution $\langle r \rangle$ approaches $\sim 0.39$. The estimate $h_{\ssm max} \approx 3.5$ for the transition point can be obtained by interpolating the crossing points of $\langle r \rangle$ for successively larger system sizes. \textbf{B and C}: $|\overbar{M(t)}|^2$ for three initial states of the environment: the groundstate (GS, orange), infinite temperature state ($T=\infty$, blue) and the N\'eel state (black). The left panel shows results for  $h_{\ssm max} = 1.0$ whereas the right panel shows results for $h_{\ssm max} = 5.0$. From the $\langle r \rangle$ statistic, we know that these are two extremes in the thermalizing and MBL regime, respectively. In the latter, the coherence is retained on much longer timescales for all of the initial states. The N\'eel state is a convenient choice of state, as it basically extracts the shape of the envelope of the decay. \textbf{D}: The real (blue) and imaginary (orange) parts of $\overbar{M(t)}$ for an initially N\'eel ordered environment with $h_{\ssm max} = 5$. The black line shows $|\overbar{M(t)}|$ while the gray bar shows $|M(t)|$ as obtained from the entropy (see main text). 
			\label{fig:figure2}}
	\end{figure*}
\end{center}

The transition is visible in the statistics of the eigenvalue distribution \cite{Meh04a, Oganesyan2007, Serbyn2015}. More precisely, the average level-spacing between eigenvalues evolves from a Wigner-Dyson distribution at low values of $h_{\ssm max}$ (in the thermalizing phase), to a Poissonian distribution in the MBL phase (see panel $\textbf{A}$ of Fig.~\ref{fig:figure2}).
The absence of thermalization is not due to fine-tuning but arises from an emergent integrability. This integrability is encoded in (quasi-) local integrals of motion (LIOM's), also referred to as l-bits (for localized)~\cite{Serbyn2013b, Ros2015, Imbrie2016, Huse2014,Rademaker2016,Chandran2015}. We will see in the remainder of this Letter that the combination of the aforementioned phenomena, i.e., violation of ETH, the change in level statistics, and the emergent integrability will largely influence the decoherence properties of the attached probe spin \cite{Lag2005, Brox2012}.

A lot of recent work in the field of interacting disordered systems has been devoted to devising observable criteria for the MBL transition. As opposed to direct measurements of the MBL system, here, we present an alternate view: we use an external spin to probe the MBL system described by Eq.~\ref{eq:environmentHamiltonian}. In particular, we show that the decoherence of the external spin provides unambiguous signatures of MBL and directly manifests the fundamental logarithmic growth~\cite{Znidaric2008,Bardarson2012a} of the entanglement entropy expected in MBL systems.

The Hamiltonian of the full system reads
\begin{equation}
		H = H_{\textrm{s}} + H_{\textrm{sp}},
\end{equation}
where $H_{\textrm{s}}$ is given in Eqn.~(\ref{eq:environmentHamiltonian}) and $H_{\textrm{sp}}$ describes the system-probe interaction. 
This interaction Hamiltonian is given by:
\begin{equation}
		H_{\textrm{sp}} = 2 \sum_i g_i S_i^z \sigma^z,
\end{equation}
where $\sigma^a$ ($a \in \{x,y,z\}$) is the spin-$1/2$ operator describing the probe spin. We do not include any intrinsic Hamiltonian for the probe spin since we are fundamentally interested in the decoherence rather than relaxation. However, a magnetic field acting on the spin along the $z$ direction induces a pure phase factor and does not play a role in what will be discussed below. We will primarily discuss the coupling of the probe spin to one end of the chain, i.e. $g_L = g$ and all other $g_i$ are set to zero. As opposed to an infinite range coupling of the spin to disordered baths, which is not particularly sensitive to localization \cite{Wino2009}, by coupling the test-spin to a single site we will be susceptible to localization and the associated emergent integrability of the MBL phase. Let us now discuss this decoherence dynamics.

\section{Dynamics}
At time $t = 0$, we assume the combined system to be in a factorizable initial density matrix $\Omega = \rho_{\textrm{s}} \otimes \rho$, with $\rho_{\textrm{s}}$ the density matrix of the system. The probe spin is initialized in a pure state $\rho = |\psi_0\rangle \langle \psi_0|$ with $|\psi_0\rangle = \alpha\mid \uparrow \rangle + \beta\mid \downarrow \rangle$. This choice of initial state and interaction terms leads to a pure decoherence of the probe spin.

Turning on the coupling at $t=0^+$, the total density matrix undergoes the usual Hamiltonian time evolution. In the basis $\{\mid\uparrow\rangle, \mid\downarrow\rangle\}$, the time evolved reduced density matrix $\rho={\rm Tr}_{\textrm{s}}\Omega$ of the probe spin can be formally expressed as
\begin{equation}
		\rho(t) = \begin{pmatrix} |\alpha|^2 & M(t) \alpha^* \beta \\ M(t)^* \alpha \beta^* & |\beta|^2 \end{pmatrix}.
\end{equation}
The factor
\begin{equation}\label{moft}
		M(t) = \text{Tr}\Big( e^{-i(H_{\textrm{s}} + g S_L^z)t} \, \rho_{\textrm{e}} \, e^{i(H_{\textrm{s}} - g S_L^z)t} \Big)
\end{equation}
is the measure of the spin coherence at time $t$. We will be mostly concerned with the disorder average of $M(t)$, which we denote with an overbar as $\overbar{M(t)}$, but let us make a few remarks on the behaviour of $M(t)$ before disorder averaging. The expression in Eq.~\eqref{moft} is closely related to the Loschmidt echo, which is used as a measure of quantum chaos \cite{Pastawski01}. It is valid for arbitrary coupling between the probe spin and the system and goes beyond the standard Born-Markov approximation used in typical studies of decoherence. 
At asymptotic times, the coherence $\overbar{M(t\to\infty)} \to 0$. Due to the finite size of the system and the purely unitary time-evolution, $M(t)$ for a single realization of disorder does not decay however. In fact, it shows recurrences that fully revive the coherence of the probe spin. The times at which these recurrences occur depend on the particular disorder realization, and hence averaging these realizations leads to the eventual decay of $\overbar{M(t)}$. However, the transient behaviour of $\overbar{M(t)}$, which shows how the probe spin decoheres, may depend on the properties of the environment. We exploit this aspect to see if the transient behaviour of the decoherence is indeed a sensitive probe of the underlying many body localization. Such disorder averaging is feasible also in experiments, although reasonably with a smaller number of averages than what we shall consider below. A small number of averages ($\sim 30$) was however observed to already highlight the relevant features~\cite{Smith2015}.

To probe MBL, we consider three different initial states for the environment: i) the groundstate, ii) the infinite temperature density matrix $\rho_s = \mathds{1}/2^L$ and iii) in analogy with experiments on MBL, a high-energy N{\'e}el ordered state~\cite{Schreiber2015}. States (ii) and (iii) are apt choices as they emphasize the MBL aspect, since eigenstate thermalization is expected to be violated for highly excited states in particular.

In order to obtain the coherence function $M(t)$ one can go along different routes. For small values of the coupling $g \ll h_{\ssm max}, J$, where the Born approximation holds, the coherence is essentially determined by the end spin correlator $\text{Tr}\left( \rho_{\textrm{s}} S_L^z(t) S_L^z(0) \right)$ \cite{Camalet2007}. For arbitrary values of the coupling, on the other hand, one can use exact numerical diagonalization methods to evaluate $M(t)$. In the following, we set the exchange coupling $J=1$, the probe coupling $g=1$, and restrict ourselves to the total $\sum_{i=1}^L S_i^z = 0$ sector. This restriction allows us to directly study the level statistics of the model. Moreover this constraint is physically relevant in experiments with cold atoms, where one works within a restricted spin sector. 

\section{Results}
We use exact diagonalization to calculate the coherence for an environment of size $L=8$, and calculate the average $\overbar{M(t)}$ over $\sim 10^6$ disorder realizations. Already for these systems sizes we observe the main features of the level statistics expected for the thermodynamic limit (see panel $\textbf{A}$ of Fig.~\ref{fig:figure2}). 

Panels $\textbf{B}$ and $\textbf{C}$ of Fig.~\ref{fig:figure2} show results for the disorder averaged coherence of the three choices of initial states, at disorder strengths deep in the delocalized ($h_{\ssm max} = 1$) and localized ($h_{\ssm max} = 5$) regimes. For all choices of the initial state, we observe starkly different behaviours of $|\overbar{M(t)}|^2$ in the two phases. The coherence decays on much shorter time scales in the delocalized regime for $ h_{\ssm max} \lesssim 3.5$ as opposed to it persisting until much longer times in the localized phases. The initial N\'eel ordered state essentially captures the envelope of the $T=\infty$ state coherence, up to a linear scaling of time by a factor two. Therefore, we focus exclusively on the N\'eel state and analyze the different time regimes.

The real and imaginary parts of $\overbar{M(t)}$ for the N\'eel state in the MBL regime, corresponding to the expectation values of the probe spin in the $x$- and $y$-direction respectively, are plotted in panel $\textbf{D}$ of Fig.~\ref{fig:figure2}. The resulting absolute value $|\overbar{M(t)}|$ is shown in black. Additionally the panel shows an approximation to $|\overbar{M(t)}|$ in gray, obtained from the entropy $S(t) = -\text{Tr} \rho(t) \ln \rho(t)$ of the probe spin. For times $t > 10$ the two curves agree. In the small $M(t)$ limit namely, $S(t)$ can be expanded to $S(t) \approx \ln 2 - \frac{1}{2}|M(t)|^2$ (we have set $\alpha = \beta = 1/\sqrt{2}$ for simplicity). 

On short timescales ($0 < t \lesssim J^{-1}$) the coherence shows a near universal Gaussian decay $|M(t)|^2 \sim e^{-(t/t^*)^2}$ for all disorder strengths, with $t^*$ having only a very weak dependence on $h_{\ssm max}$ (not shown). This is expected from a small $t$ expansion of the expression for $M(t)$ in Eq.~\ref{moft}, in which the linear term in $t$ is imaginary.

Substantial deviations arise at intermediate and longer times. The slower decay of coherence in the MBL phase for $h_{\textrm{max}} \gtrsim 3.5$ can be explained by considering the properties of the environment. In the large disorder limit where the level statistics shows Poissonian behaviour, an extensive number of local conserved integrals of motion (LIOMs) are postulated to exist~\cite{Serbyn2013b, Ros2015, Imbrie2016, Huse2014,Rademaker2016,Chandran2015}. In particular, if the probe spin couples dominantly to one LIOM (in this case, at the end of the chain), any diffusion of information to the other end will require times exponential in the system size. This implies that the initial coherence of the probe spin is preserved longer since it effectively Rabi-oscillates with this LIOM at the end, ensuring that even after averaging over disorder the coherence is retained.
 
\begin{center}
	\begin{figure}[t!]
			\includegraphics{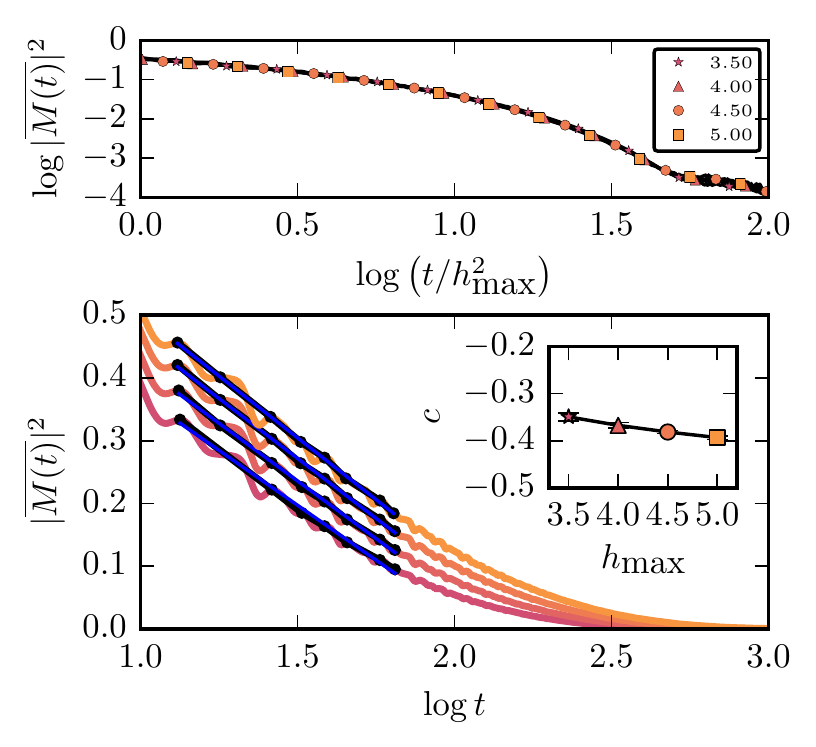}
			\caption{\textbf{Top panel}: In the MBL regime for $h_{\ssm max} \gtrsim 3$ the curves for $\log \overbar{|M(t)|}^2$ can be collapsed by rescaling $t \to t/h_{\ssm max}^2$ (shown are the curves for $h_{\ssm max} = 3.5, 4.0, 4.5 and 5.0$). This scaling works well in the intermediate timescale. The curves for smaller $h_{\ssm max}$ are not shown. Various goodness-of-fit parameters indicate that for those values of $h_{\ssm max}$ a power-law decay fits better than a logarithmic one, and hence the scaling-behaviour of those curves deviates. \textbf{Bottom panel}: The intermediate times can be fit well by a logarithmic decay, $|\overbar{M(t)}|^2 \sim c \log t$, with $c \approx -0.37$ (shown in the inset as a function of $h_{\ssm max}$). This is in agreement with the observed logarithmic growth of the entanglement entropy in the MBL phase. 
			\label{fig:logfit} }
	\end{figure}
\end{center}

For intermediate times $J^{-1} \ll t \ll t_f(h_{\ssm max})$ during which most of the decoherence occurs, $|\overbar{M(t)}|$ manifests a scaling behaviour for $h_{\ssm max} \gtrsim 3.5$ while no such scaling is seen for $h_{\ssm max} \lesssim 3.5$. Specifically, $M(t)$ and the entropy $S(t)$ for different $h_{\ssm max}$ in the localized regime can be collapsed onto a single curve by rescaling the time axis $t$ to $t/h_{\ssm max}^\alpha$, with a disorder independent exponent $\alpha = 2$ as shown in the top panel of Fig~\ref{fig:logfit}. 

An important signature of MBL is the logarithmic growth in time of the bipartite entanglement entropy~\cite{Znidaric2008,Bardarson2012a}. Here, as shown in Fig.~\ref{fig:logfit}, in this transient time regime, the coherence $\vert \overbar{M(t)} \vert^2 \propto c \log (t)$, implying that the entropy of the spin grows as $S(t)\propto -c \log (t)$. Concomitant with the scaling behaviour of $|\overbar{M(t)}|$ discussed above, we find that the coefficient $c \approx -0.37$ varies only weakly with disorder strength. Since the scaling implies $t_f \propto h_{\ssm max}^2$, the logarithmic regime lasts for longer times as the disorder strength increases. This result carries some importance, as it shows that the decoherence of the probe spin, which is relatively easy to measure, experimentally provides a direct way of measuring the logarithmic entanglement growth.

\begin{center}
	\begin{figure}[t!]
			\includegraphics{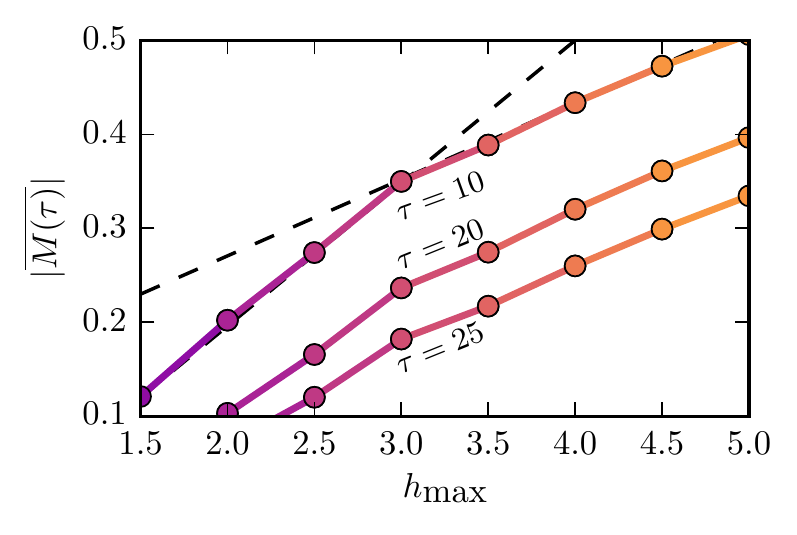}
			\caption{The values of $\overbar{|M(\tau)|}$ at fixed times $\tau = 10,20$ and $25$ in the intermediate time regime, as a function of disorder strength $h_{\ssm max}$. The dependence of $\overbar{|M(\tau)|}$ on $h_{\ssm max}$ shows a change of slope for the localized regime. The black lines are guides for the eye. 
			\label{fig:MofTau} }
	\end{figure}
\end{center}

The difference in behaviour for the delocalized and MBL regimes can be further exemplified by looking at the coherence at representative times $\tau$ in this intermediate regime. Fig.~\ref{fig:MofTau} shows the coherence values for $\tau = 10,20$ and $25$ as a function of disorder strength, indicating a change of slope in the way these times evolve at $h_{\ssm max} \approx 3$. The slopes of $\overbar{M(\tau)}$ versus disorder strength in the MBL regime are identical, but vary strongly in the delocalized regime (not shown). 

Lastly, we mention that at asymptotic times $t \gg t_f(h_{\ssm max})$ a new regime sets in, where $ \ln |M|(t) \propto - t^\nu$ where $\nu \approx 4$ and has a slight dependence on $h_{\ssm max}$ for the MBL regime.

\vspace{-1.5cm}
\section{Discussion}
The curves $|\overbar{M(t)}|$ in the MBL regime for different coupling strength $g$ are simply related by a scaling of the time $t \to gt$. Consequently, all of the features of the coherence in the MBL regime discussed earlier can be observed for weak as well as strong coupling of the probe to the spin chain. This highlights the fact that in the MBL phase, the states are localized and the strength of the coupling $g$ only determines that time scale over which this state is probed. At smaller $h_{\ssm max}$, the system is delocalized and no such scaling holds. We have checked that our conclusions hold even if the probe couples to a spin in the middle of the chain as well as larger system sizes. By studying chains with $10$, $12$ and $14$ sites we have confirmed that (due to the LIOMs) there is little to no effect of scaling in system size.

To conclude, we showed that decoherence of a single spin coupled to a MBL bath is a valuable probe of many body localization. We established that in the MBL phase the coherence shows scaling behaviour with respect to the strength of disorder $h_{\ssm max}$ as well as with the spin-system coupling constant $g$. The predicted logarithmic growth of the bipartite entanglement entropy in the MBL phase manifests itself directly in the logarithmic growth of the entropy or, equivalently, in a logarithmic decay of the coherence of the probe spin in the intermediate time regime. From the real time decay of the coherence, we extract clear measures which differentiate between the delocalized and MBL regimes of the environment. Our setup may possibly be realized in cold atom systems or more analogously in a trapped-ion system, where one has a high degree of single spin control. The real-time evolution of the coherence of the probe quit for each realization of disorder can be measured in trapped ion systems using spin fluorescence.

\section*{Acknowledgements}
We acknowledge financial support from the Swiss National Science Foundation (SNSF). 

\appendix

%


\end{document}